\documentclass[12pt]{iopart}
\usepackage{graphicx}

\begin{document}

\title[]{Primordial fluctuations and cosmological inflation
       after WMAP 1.0}
\author{Dominik J.~Schwarz\dag\ and
        C\'{e}sar A.~Terrero-Escalante\dag\ddag}
\address{\dag\ Department of Physics, CERN, Theory Division, 1211 Geneva 23,
         Switzerland}
\address{\ddag\ Departamento de F\i\'sica, Centro de Investigaci\'{o}n y
         Estudios Avanzados del IPN, Apdo.~Postal 14-740, 07000 M\'{e}xico D.F.,
         M\'{e}xico}
\eads{\mailto{Dominik.Schwarz@cern.ch},\ \mailto{Cesar.Terrero@cern.ch}}

\begin{abstract}
The observational constraints on the primordial power spectrum have tightened
considerably with the release of the first year analysis of the WMAP
observations, especially when combined with the results from other CMB
experiments and galaxy redshift surveys. These observations allow us
to constrain the physics of cosmological inflation:
\begin{enumerate}
\item The data show that the Hubble distance is almost constant during
inflation. While observable modes cross the Hubble scale, it changes by less
than $3\%$ during one e-folding: $\dot{d}_{\rm H} < 0.032$ at $2\sigma$.
The distance scale of inflation itself remains poorly constrained:
$1.2 \times 10^{-28} \mbox{\ cm} < d_{\rm H} < 1 \mbox{\ cm}$.
\item We present a new classification of single-field inflationary scenarios
(including scenarios beyond slow-roll inflation), based on physical criteria,
namely the behaviour of the kinetic and total energy densities of the inflaton
field. The current data show no preference for any of the scenarios.
\item For the first time the slow-roll assumption could be dropped from the
data analysis and replaced by the more general assumption that the Hubble scale
is (almost) constant during the observable part of inflation.
We present simple analytic expressions for the scalar and tensor power spectra
for this very general class of inflation models and test their accuracy.
\end{enumerate}
\end{abstract}

\begin{flushleft}
\end{flushleft}

\maketitle

\section{Introduction and results}

With the release of the analysis of the first year WMAP data
\cite{Bennett:2003bz,Spergel:2003cb} and the first 3d SDSS power
spectrum \cite{Tegmark:2003uf}, we got splendid confirmation of the
long-standing expectation that the primordial power spectrum of
density fluctuations is almost scale-invariant, as anticipated
already by Harrison and Zel'dovich in the 1970s. The simplest and
most elegant mechanism known to produce primordial spectra with
that property is cosmological inflation. Its beauty lies in the fact that
we do not need to invent untested physical principles: quantum
mechanics and general relativity in four space-time dimensions are
good enough. What is needed is a cosmic substratum that gives rise to an
epoch of accelerated expansion of the very early Universe. A
positive energy density of the vacuum is the simplest example. Trying to figure
out the underlying mechanism leads to the pressing question: {\it What is the
scale of cosmological inflation?} It is probably the most poorly constrained
energy scale in cosmology; we could argue that cosmological inflation
happens somewhere between the electroweak scale (requiring a mechanism to
create baryons after the end of inflation) and the highest possible energy
scale, which might be the scale of quantum gravity. This implies an
uncertainty of some $16$ orders of magnitude.

Although we have no clue on how the world looks like at scales beyond
the standard model of particle physics, we can make some generic
predictions about the fluctuations in the energy density and
space-time structure that are seeded by quantum fluctuations
during cosmological inflation \cite{Mukhanov:xt,Starobinsky:ty}.

We assume that the first prediction of cosmological inflation, spatial
flatness of the Universe, is well established and take $\Omega = 1$ (see
\cite{Benoit:2002mm,Spergel:2003cb} for a detailed discussion). Further, we
assume that the observed perturbations are of isentropic nature, which is
another prediction of cosmological inflation, unless there is some matter
component in the Universe that never ever coupled to the radiation fluid.
The most convincing evidence for the dominance of isentropic fluctuations
comes from the recent detection of polarization in the cosmic microwave
background (CMB) \cite{Kovac:2002fg,Kogut:2003et}, which requires the presence
of a quadrupole anisotropy at the moment of photon decoupling and therefore
shows that large-scale fluctuations existed back then. In the following we
focus on the power spectra of scalar and tensor fluctuations, which contain all
the information if, again as predicted by inflation, the fluctuations are
Gaussian (which is consistent with the data \cite{Komatsu:2003fd}).

Two important inputs are required to predict the inflationary
power spectra: the Hubble rate $H$ during cosmological inflation
as a function of the scale factor $a$ (or, for later convenience, as
a function of the logarithm of the scale factor, denoted by $N$) and
the number of dynamical degrees of freedom that drive inflation.
In the case of de Sitter space-time this number is zero, but in
such a model inflation does not end; the vacuum energy thus must
be dynamical. In the simplest scenarios with an exit from
inflation a single scalar field $\varphi$, usually called {\it
inflaton}, is responsible for the acceleration of the Universe. We
will restrict the following considerations to that case, with
equations of motion,
\begin{eqnarray}
\label{eq:KGphi}
\ddot{\varphi} + 3H\dot{\varphi} + V'(\varphi) = 0 \, ,\\
\label{eq:Feq}
H^2 = \frac{1}{3}\left[\frac{\dot{\varphi}^2}{2}+V(\varphi)\right]\, .
\end{eqnarray}
Here, $V(\varphi)$ is the potential energy density of the inflaton field,
the dot stands for a derivative with respect to cosmic time $t$,
and the prime denotes a derivative with respect to the inflaton field.
We use Planck units, $8\pi G = c = \hbar = 1$.

In principle, observations should allow the reconstruction of $H(N)$, but
since only a finite interval of wave numbers $k$ is accessible to
observations, we can only probe a finite and rather short interval
$\Delta N$. An efficient encoding of $H(N)$ is by the so-called
horizon-flow functions, evaluated at a pivot point $N_*$ adapted
to the experimental set-up. The horizon-flow functions are a
generalization of the slow-roll parameters \cite{Liddle:1994dx}
and are defined recursively
as the logarithmic derivatives of the Hubble scale with respect to
the number of e-foldings $N$ \cite{Schwarz:2001vv}:
\begin{equation}
\label{eq:hff}
\epsilon_{m+1} \equiv {{\rm d} \ln |\epsilon_m|\over {\rm d} N}
\quad \forall m \geq 0, \quad
\epsilon_0 = {H(N_{\rm i})\over H(N)}\, ,
\end{equation}
where $N_{\rm i}$ denotes an arbitrary `initial' moment. The necessary
condition for inflation to take place ($\ddot{a} > 0$) becomes
$\epsilon_1<1$, and we assume here that the weak energy condition and null
energy condition hold true, i.e.~$\epsilon_1 \geq 0$.
We define the {\it graceful exit} from inflation as the moment when
$\epsilon_1$ crosses unity. Specifying the set
$\{\epsilon_m(N_*)\}$ is equivalent to specifying $H(N)$. A truncation of
the set $\{\epsilon_m(N_*)\}$ corresponds to an incomplete knowledge of
the evolution of the Hubble rate.

In single field slow-roll models of inflation, all the horizon-flow
functions are typically small and, hence, equivalent to the
slow-roll parameters. In terms of the inflaton potential
$V(\varphi)$ and its derivatives with respect to the inflaton
field, the first two flow functions are given by
\cite{Leach:2002ar}:
\begin{equation}
\epsilon_1 \approx
\frac{1}{2}\left(\frac{V^\prime}{V}\right)^2\, , \qquad
\epsilon_2 \approx 2
\left[\left(\frac{V^\prime}{V}\right)^2-\frac{V^{\prime\prime}}{V}\right]\, .
\end{equation}

CMB experiments (the most recent
results come from CBI \cite{Pearson:2002tr,Readhead:2004gy}, Archeops
\cite{Benoit:2002mk}, ACBAR \cite{Kuo:2002ua}, VSA
\cite{Grainge:2002da,Dickinson:2004yr} and WMAP \cite{Bennett:2003bz}) and
galaxy redshift surveys (especially 2dF \cite{Percival:2001hw,Tegmark:2001jh}
and SDSS \cite{Tegmark:2003uf}) have measured the amplitude $A$ and the
spectral index $n$ of density fluctuations and constrained the
running\footnote{Two recent works on new data from CBI
\cite{Readhead:2004gy} and VSA \cite{Rebolo:2004vp} claim to see evidence for
a large (compared to expectations from cosmological inflation) running of the
spectral index with negative sign, but point out a calibration issue that
weakens the evidence.}
of the spectral index ${\rm d}n/{\rm d}\ln k$, as well as
the amount of gravitational waves, typically expressed as the
tensor-to-scalar ratio $r$, that can contribute to the CMB signal.
This set of observables $\{A,n,{\rm d}n/{\rm d}\ln k,r\}$ could,
in principle, provide a measurement of the inflationary parameters
$\{H,\epsilon_1, \epsilon_2,\epsilon_3\}$.

\begin{figure}
\label{fig:fig1}
\centerline{\includegraphics[angle=270,width=0.95\linewidth]{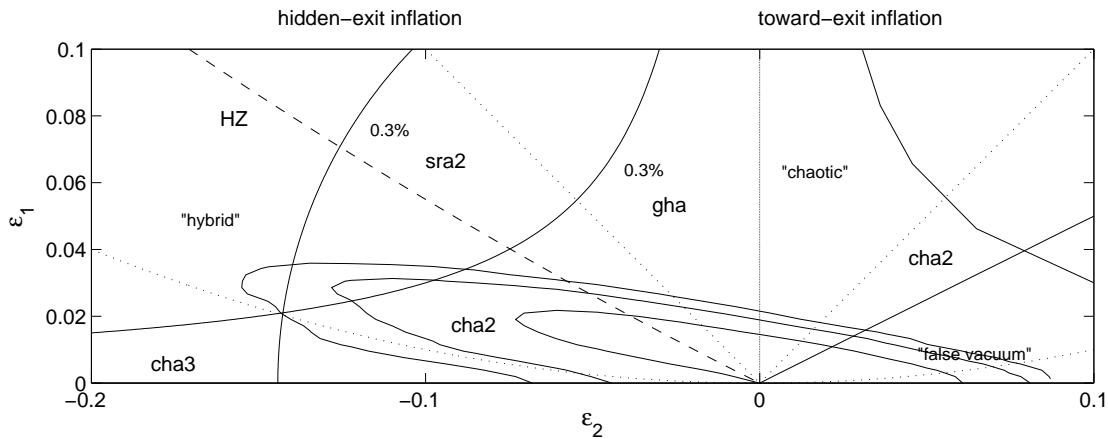}}
\caption{Observational constraints, interpretation, and accuracy of
various approximations in the $\epsilon_1$--$\epsilon_2$ plane: the
likelihood curves have been provided by S.~M.~Leach
\cite{Leach:2003us} and enclose the $1\sigma$-, $2\sigma$- and
$3\sigma$-allowed regions. Models with a scale-invariant spectrum
fall on the dashed line (HZ). An observation of $\epsilon_2 > 0$
would favour simple models of inflation, in which the evolution
toward a graceful exit is seen (toward-exit inflation). If
$\epsilon_2 \leq 0$, a more contrived scenario must be realized,
since in that case the mechanism of exit is hidden from observation
(hidden-exit inflation). The region with $\epsilon_2 > 0$ is further
split up by the (full straight) line $\epsilon_1 = \epsilon_2/2$,
which distinguishes models in which the kinetic energy density
decreases (above the line) or increases with time. The latter
requires a mechanism to ensure that the kinetic energy density is
very small initially. The solid arc centered at the origin encloses
the region in which the slow-roll approximation at second order
(sra2) provides a prediction of the amplitude at the pivot scale
better than $0.3\%$. Below the hyperbolic lines the same accuracy is
achieved for the constant-horizon approximation (cha2 at second
order, cha3 at third order) and the growing horizon approximation
(gha). The dotted lines represent the boundaries between cha3 and
cha2, as well as between cha2 and gha.}
\end{figure}

In a number of works, various combinations of experiments and
methods have been used to determine a subset of these parameters,
see especially
Refs.~\cite{Peiris:2003ff,Barger:2003ym,Kinney:2003uw,Leach:2003us,
Tegmark:2003ud,Rebolo:2004vp}.
Most interesting are the constraints in the
$\epsilon_1$--$\epsilon_2$ plane; see Fig.~\ref{fig:fig1}. The
common finding is that models with almost scale-invariant spectra
provide acceptable fits and $\epsilon_1 < 0.032$ (at $2 \sigma$)
\cite{Leach:2003us}. The
corresponding likelihood contours from that work are shown in
Fig.~\ref{fig:fig1} where the dashed line denotes
scale-invariance, up to third order corrections in the horizon-flow
functions. Thus, observations show for the first time that
$\epsilon_1 \ll 1$. The limits on $\epsilon_2$ are much less
restrictive; the situation $|\epsilon_2| > \epsilon_1$ covers a
large part of the allowed parameter space. This is due to the
so-called tensor degeneracy \cite{Efstathiou:2001cv}, i.e. one can
compensate an increase of the tensor contribution by making the
scalar spectral index bluer. Since there exist only upper limits
on the contribution from gravitational waves, the scale of
inflation cannot be fixed by present observations. However, the
combination of the upper limit on $\epsilon_1$ and the measurement
of the amplitude of scalar fluctuations also allows an
upper limit on the energy scale of inflation $V^{1/4} < 0.01
(\equiv 2.7 \times 10^{16} \mbox{\ GeV})$ \cite{Leach:2003us} or a
lower limit on the distance scale during inflation $d_{\rm H} >
1.2 \times 10^{-28} \mbox{\ cm}$. A lower (upper) limit on the
energy (distance) scale can be obtained from the facts that the
Universe contains baryons and that there is no known mechanism to
produce baryons below the electroweak scale ($100$ GeV,
respectively $1$ cm) \cite{Schwarz:2003du}. Thus observations put
the scale of inflation at least two orders of magnitude below the
Planck scale [this limit will improve proportionally to the limits on
$\sqrt{\epsilon_1}$ and will be tightened as the value of the optical
depth $\tau$ is pinned down more precisely, since
$A \exp(-2\tau)$ is the observed quantity].

Based on the observational constraints on the first two horizon-flow functions,
we do not restrict our considerations to slow-roll models in this paper.
We relax the slow-roll conditions ($\epsilon_m \ll 1$ for all $m$) to
$\epsilon_m \leq 1$ for some $m$. In Section~\ref{sec:hubble} we demonstrate
that the condition $\epsilon_1 \ll 1$ is sufficient to guarantee that the
Hubble scale is almost constant during the inflationary epoch of interest.

The physical meaning of the first and second horizon-flow functions is
discussed in Section~\ref{sec:classification} in the framework of single-field
models, and a new physical classification of inflationary scenarios is
introduced, based on the behaviour of the kinetic and total energy densities
of the inflaton. To our surprise this approach closely resembles the small
field/large field/hybrid classification of Dodelson et
al.~\cite{Dodelson:1997hr}, but is not restricted to slow-roll models.

In Section~4 we derive a procedure to self-consistently calculate
higher-order corrections at any scale $k$, given an analytic
expression of the power at the pivot scale $k_*$ as a function of
the horizon-flow functions. A similar relation has been used in
the literature to calculate the spectral index and its running so
far, but, to our knowledge, it has never been justified rigorously.

Starting from the observation that $\epsilon_1 \ll 1$,
we show that the slow-roll approximation is no longer needed in
the analysis of the data and that it could be replaced by a more
efficient approximation (constant-horizon approximation
\cite{Schwarz:2001vv}). The latter is more efficient in the sense
that we need to know less horizon-flow functions in order to
predict the power spectra to a given accuracy, as compared with
the slow-roll approximation.

We map the classification of models from the $\epsilon_1$--$\epsilon_2$
plane to the space spanned by the tilt of the
spectrum,
$n-1$, and the ratio of tensor and scalar amplitudes $r$ in
Section~5. At leading order in the horizon-flow functions this is a
one-to-one mapping; including higher orders introduces an ambiguity,
due to the running of the spectral index. We thus argue that the
$\epsilon_1$--$\epsilon_2$ plane is more fundamental. For practical
purposes, the $r$--tilt plane is nevertheless a useful tool.

In Section~6 we give examples to demonstrate the accuracy
of the new method and discuss the applicability of the new
method in the light of the recent data.

Our main results are summarized in Fig.~\ref{fig:fig1}. We overlay
our new classification of models on the likelihood contours as
obtained by Leach and Liddle \cite{Leach:2003us}. We also indicate
which approximation to the power spectrum at the pivot scale is
best suited for which region of parameter space. We assume an
accuracy goal of $0.3\%$ at the pivot, which puts the theoretical
errors safely away from the systematic and statistical errors of WMAP
and Planck.

\section{Hubble horizon flow during inflation}
\label{sec:hubble}

The Hubble horizon, $d_{\rm H} \equiv c/H$ (recall that $c=1$),
is roughly the size of the region where causal processes can
take place during one Hubble time $1/H$. An inflationary epoch is
characterized by a decreasing comoving Hubble horizon, $d_{\rm
H}/a$. Constant vacuum energy is the simplest (but unphysical, since
inflation would continue forever) kind of matter leading to an
inflationary universe. In this scenario, the Hubble scale is
constant. More generally, during inflation, $d_H$ is expected to
vary slowly during a given number of e-foldings $N$: ${\rm
d}N/{\rm d}t=H$. The behaviour of the Hubble horizon can be
parameterized as
\begin{eqnarray}
d_{\rm H}(N)&=& d_{\rm H}(N_*)\left[1+\epsilon_1(N-N_*)+ \frac 1
2(\epsilon_1^2+\epsilon_1\epsilon_2)(N-N_*)^2
 + \cdots \right]\, ,
\end{eqnarray}
where the coefficients in this expansion are expressed in terms of
the horizon-flow functions $\epsilon_m$ at the pivot point. One
can see that, if $\epsilon_1 \ll 1$ and $\epsilon_m<1$
for any $m>1$, then $d_{\rm H} \approx$ constant. For such cases,
the smaller $\epsilon_1$ is, the larger the other horizon-flow
functions can be, since $\epsilon_1$ is the leading factor of all
coefficients of the higher-order terms in the series. We can use
this observation as the starting point for what we call the
constant-horizon approximation ($\epsilon_1 \ll 1$ and $\epsilon_m
<1$ for all $m>1$). As is seen in Fig.~\ref{fig:fig1} and
discussed in the introduction, the data show that $\dot{d}_{\rm H}
\equiv \epsilon_1 \ll 1$.

\section{Classification of single scalar field inflationary models}
\label{sec:classification}

The number of proposed models of cosmological inflation is very large and
keeps growing. Current observations allow us to constrain the number of
successful models, and future observations are expected to play a stronger
role in that direction. For a more effective use of the data analysis, it is
useful to group the inflationary scenarios according to some commonly
applicable criteria, i.e.~independent of parameters such as masses or coupling
constants, which are strongly model-dependent. For slow-roll inflation, such
a classification was suggested by Dodelson, Kinney and Kolb
\cite{Dodelson:1997hr}, based on the shape of the inflaton potential
(and expressed as conditions on the slow-roll parameters).

Here we present a new scheme in terms of the horizon-flow functions, which
closely resembles the Dodelson et al.\ classification, but avoids misleading
terminology, allows the inclusion of models beyond the slow-roll class,
and is based on physical criteria on the kinetic and total energy densities.

The idea is to investigate how the kinetic and potential energy of the
inflaton field change with time, both absolute and relative to each other.
It is thus useful to rewrite the first two horizon-flow functions as
\begin{eqnarray}
\label{eq:e1energy}
\epsilon_1 = 3{\dot{\varphi}^2/2\over \dot{\varphi}^2/2 + V}\, ,\\
\label{eq:e2energy}
\epsilon_2 = 2\left({\ddot{\varphi}\over H\dot{\varphi}} +
\epsilon_1\right)\, ,
\end{eqnarray}
where we used Eqs.~(\ref{eq:KGphi}) and (\ref{eq:Feq}).
Using Eqs.~(\ref{eq:KGphi}) and (\ref{eq:e2energy}),
the time variation of the potential energy is found to be
\begin{equation}
\label{eq:Vdot}
\dot{V} = - H \dot{\varphi}^2 \left(3 - \epsilon_1 +
\frac{\epsilon_2}{2}\right).
\end{equation}
Since $H \dot{\varphi}^2 > 0$, the potential energy density can never increase
for physically meaningful values of $\epsilon_1$ and $\epsilon_2$. Thus, we
focus on the behaviour of the kinetic energy density.

According to Eq.~(\ref{eq:e1energy}),
$\epsilon_1/3 \geq 0$ measures the ratio of kinetic energy density to total
energy density.  Using definition (\ref{eq:hff}),
we can ask how this ratio changes with time during inflation:
\begin{equation}
\frac{\rm d}{{\rm d} t}\frac{\epsilon_1}{3} = H \frac{\epsilon_1}{3}\epsilon_2.
\end{equation}
Since $H$ and $\epsilon_1$ are positive, $\epsilon_2 = 0$ marks a
borderline between two physically different cases:
increasing ($\epsilon_2 > 0$)
and decreasing ($\epsilon_2 < 0$) kinetic energy density
with respect to the total energy density of the inflaton.

For a more complete understanding of the inflationary dynamics, we
must also look at the absolute time variation of the kinetic
energy density ${\rm d}{(\dot{\varphi}^2/2)}/{\rm d}t$, which,
using Eqs.~(\ref{eq:KGphi}) and (\ref{eq:Vdot}), can be written as
\begin{equation}
\dot{\varphi}\ddot{\varphi} = H \dot{\varphi}^2
\left(\frac{\epsilon_2}{2} - \epsilon_1\right).
\end{equation}
For $\epsilon_1 > 0$ we have $\dot{\varphi}\neq 0$ and thus
$\epsilon_2 = 2 \epsilon_1$ is another borderline between
two different physical scenarios in which kinetic energy density grows
($\epsilon_2 > 2 \epsilon_1$) or falls ($\epsilon_2 < 2 \epsilon_1$) with time.
We thus arrive at the following classification:
\begin{itemize}
\item $\epsilon_2 > 0$: the kinetic energy density is increasing with respect
to the total energy density. This is a necessary condition for evolving
toward a graceful exit of inflation. Thus, such models could be called
{\it toward-exit} models, since it can be argued that the approach to a
graceful exit of inflation is observed in that case (although a more
complicated interpretation remains possible).

There are two subcases, corresponding to whether or not the kinetic
energy density grows with time. It would be expected that this provides a
criterion to discriminate between the false vacuum models and chaotic models
in the slow-roll phase, since false vacuum models start out with vanishing
kinetic energy density, whereas chaotic models always have a non-vanishing
kinetic energy density.
\begin{itemize}
\item $\epsilon_2 > 2 \epsilon_1$: kinetic energy density grows with time.
An example of this are false vacuum models arising, e.g.~in superstrings models,
with potential $V = V_0 - m^2 \varphi^2/2$ \cite{Binetruy:ss}. Here, in
the slow-roll phase, $\epsilon_2 \approx 4 (V_0/m^2 \varphi^2) \epsilon_1$,
and the condition is met as long as $V_0 > m^2\varphi^2/2$. False vacuum
models need to provide a mechanism that gives rise to {\it small kinetic
energy} density of the inflaton before observable modes cross the Hubble
scale during inflation.

\item $\epsilon_2 < 2 \epsilon_1$: kinetic energy density is decreasing
with time. Monomial potentials $V = \lambda \varphi^n/n$ with chaotic initial
conditions give rise to models of that kind. During their slow-roll
phase $\epsilon_2 \approx (4/n) \epsilon_1$, and the criterion is met
for $n > 2$. Here no extra mechanism is needed to tune the kinetic energy
density initially, and one could argue that these are the {\it simplest}
models.

\item $\epsilon_2 = 2 \epsilon_1$: kinetic energy density is constant.
For slow-roll models, this is realised for chaotic inflation with a
quadratic potential ($n=2$) \cite{Linde:fd}.

Note that this criterion is different from the one given by Dodelson et al.\
\cite{Dodelson:1997hr} to distinguish between small- and large-field models.
In our notation and in the slow-roll approximation their borderline is
$\epsilon_2 \simeq 4 \epsilon_1$. This is the case of a linear potential
($n=1$). Naively one would think that these models should fall into the
same class as chaotic inflationary models, but actually they do not give rise
to a successful scenario, since for the linear potential
$V=V_0(1+\varphi/\varphi_0)$ we find during slow-roll $\epsilon_1 \approx
\epsilon_2/4 \approx 1/(2\varphi_0^2)$, which is constant and thus inflation
never ends.
\end{itemize}

\item $\epsilon_2 < 0$: kinetic energy density is decreasing absolutely and
relatively. Here, in order to reach a graceful exit, $\epsilon_2$ has
to change sign at some point. Thus, negative values of $\epsilon_2$ must
correspond to models in which inflation still has to go through a transition
to either another stage of inflation or to directly to stop it by
some unknown mechanism. The actual mechanism of exit is out of sight of
observations, so we could call these models {\it hidden-exit} models.

An example is the hybrid model, with an effective potential $V =
V_0 + m^2 \varphi^2/2$ in the slow-roll phase \cite{Linde:1991km}.
In that case $\epsilon_2 \approx - 4 (V_0/m^2\varphi^2) \epsilon_1$. In
order to end inflation $V_0$ must depend on yet another field which
finally drives $V_0$ to zero.

\item $\epsilon_2 = 0$: the ratio of kinetic to total energy density
is constant. If additionally all $\epsilon_m = 0$, then this is de
Sitter space-time. If only $\epsilon_1 \neq 0$, then this is
power-law inflation \cite{Lucchin:1984yf}. There also exists a number of
models where $\epsilon_2$ asymptotically converges to zero during
inflation, so that they are observationally indistinguishable from
power-law inflation (see \cite{Terrero-Escalante:2002sd} for a discussion
and further references). In all these cases there is {\it no exit} from
inflation but the models are not ruled out by current cosmological data.

Provided the slow-roll approximation holds, $\epsilon_2 = 0$ coincides with
the limit between hybrid models and large-field models in the Dodelson et al.\
classification.
\end{itemize}

Summarizing, our three classes are
i) hidden-exit inflation ($\epsilon_2 \leq 0$),
ii) toward-exit inflation with general (``chaotic'') initial conditions
($0 < \epsilon_2 \leq 2 \epsilon_1$), and
iii) toward-exit inflation with special (e.g.~false vacuum) initial
conditions ($0 < 2\epsilon_1 < \epsilon_2$).

At this point it is convenient to note that the here introduced by
us classification involves only exact expressions. We confront this
new classification with the observational constraints in
Fig.~\ref{fig:fig1}. Although the biggest piece of allowed parameter
space falls into the hidden-exit inflation class, this cannot be
seen as a preference of the data for this scenario, since the shape
of the likelihood curves is due to the tensor degeneracy. At present
all scenarios are consistent with the data. A better determination
of the spectral index (lines parallel to the dashed line correspond
roughly to fixed values of the spectral index) could in principle
rule out some of the possibilities; there is thus hope to learn more
about inflation, well before we can expect to get a hand on the
tensor contribution via observations of the B-polarization pattern
of the CMB.

\section{Power spectra}

The accurate prediction of inflationary perturbations have been of
concern since Mukhanov and Chibisov \cite{Mukhanov:xt} realized that
density and space-time perturbations during inflation could be the seeds
for large-scale structure formation. The calculation of inflationary
power spectra requires the solution of the mode equations for scalar and
tensor fluctuations. The assumption that scalar and tensor perturbations are
quantum fluctuations of the vacuum originally fixes the power spectra uniquely.
The mode equations are of the same type as the Schroedinger equation, which
happens to be difficult to solve, even for the simple models (see
\cite{Stewart:1993bc,Wang:1997cw,Martin:1999wa,Gong:2001he,Schwarz:2001vv,
Leach:2002ar,Tsamis:2003px} for details and references).

It proved useful to expand the primordial power spectrum in a
Taylor series around a pivot scale $k_* = (aH)(N_*)$.
Within a time interval $\Delta t \sim 1/H$, the modes in the
logarithmic frequency interval $\Delta \ln k = \ln(e aH) - \ln(aH)
= 1$ cross the Hubble scale\footnote{The `crossing' of the Hubble
horizon is defined here to take place when $k = aH$.}. It is thus
natural to expand in terms of $\ln (k/k_*)$:
\begin{equation}
\label{eq:power}
{\cal P}(k) = \tilde{\cal P}(k_*) \sum_{n\geq 0} {a_n\over n!}
\ln^n\left(\textstyle{k\over k_*}\right),
\end{equation}
where the coefficients are defined by
\begin{equation}
a_n(k_*) = {{\rm d}^n \over {\rm d} \ln^n k}
\left. {{\cal P}(k)\over \tilde{\cal P}(k_*)}\right|_{k = k_*}.
\end{equation}
These coefficients depend on the horizon-flow parameters in such a
way that they are regular in the limit $\epsilon_m \to 0$; in fact
$a_0 \to 1$ and $a_n \to 0$ for all $n > 0$. Now it becomes
obvious why we have separated a factor that represents the leading-order
prediction for the amplitude at the pivot scale
\begin{equation}
\tilde{\cal P}_{\rm S}(k_*) \equiv {H_*^2 \over 8 \pi^2
\epsilon_1}
\end{equation}
for scalar perturbations, and
\begin{equation}
\tilde{\cal P}_{\rm T}(k_*) \equiv {2 H_*^2 \over \pi^2}
\end{equation}
for tensor perturbations. In the notation of the WMAP team,
${\cal P}_{\rm S}(k_*) \equiv 10^{-9} A$ and
${\cal P}_{\rm T}(k_*)/{\cal P}_{\rm S}(k_*) \equiv r$.

\subsection{Independence from the pivot point}

The physical power spectrum must not depend on the arbitrary choice of a
pivot scale $k_*$ in the above expansion:
\begin{equation}
{{\rm d} {\cal P}(k)\over {\rm d} \ln k_*} \equiv 0 \, .
\end{equation}
Evaluation of this expression provides us with the non-trivial relation
\begin{equation}
\sum_{n \geq 0}{1\over n!}\ln^n \left({k\over k_*}\right)
\left[{{\rm d} \ln \tilde{\cal P}\over {\rm d} \ln k_*} a_n
+ {{\rm d}a_n\over {\rm d} \ln k_*}\right]
- \sum_{n\geq 1} {a_n \over (n-1)!}\ln^{n-1}\left({k\over k_*}\right) = 0
\, .
\end{equation}
A comparison of the coefficients of $\ln^n (k/k_*)$ finally leads to
a recursion relation for the higher coefficients of the Taylor series:
\begin{equation}
\label{eq:an}
a_{n+1} = {1\over 1 - \epsilon_1} \left[{{\rm d}a_n\over {\rm d} N_*}
+ {{\rm d} \ln \tilde{\cal P}\over {\rm d} \ln N_*} a_n \right] \quad
\forall n \geq 0\, ,
\end{equation}
where in the last step we used the identity
\begin{equation}
{\rm d}\ln k_* = (1 - \epsilon_1) {\rm d}N_* \, .
\end{equation}
For scalars ${\rm d}\ln\tilde{\cal P}_{\rm S}/{\rm d}\ln N_* =
- 2\epsilon_1 - \epsilon_2$, whereas for tensors
${\rm d}\ln\tilde{\cal P}_{\rm T}/{\rm d}\ln N_* = - 2\epsilon_1$.
Let us note that the recursion relation (\ref{eq:an}) is an exact result:
no approximation has been made, apart from the assumption that the linear
perturbation analysis is justified.

\subsection{Approximation schemes}

Up to date, three approximation schemes have been proposed to
allow for a high-precision calculation of the spectra amplitudes.
Using recursion (\ref{eq:an}), all of these approximations yield
expressions for the coefficients $a_n$ as expansions in terms of
the horizon-flow functions. Keeping terms up to order $\epsilon^q$ (which
stands here for any monomial of $\epsilon_m$'s at order $q$) in $a_0$
allows us to calculate all terms up to order $\epsilon^{q+n}$ in $a_n$.
We will denote the order of a given approximation by the highest order
in the seed $a_0$. Besides the order $q$, a second choice that must be
made prior to data analysis is how many coefficients $a_n$ should
be included in the analysis. The common practice is that only the terms
$a_0$ and $a_1$ are taken into account, $a_2$ being included when the running
of the spectral index is included.

\subsubsection{Slow-roll approximation.}

Assuming all horizon-flow functions to be small (without assuming any hierarchy
among them), the equation of modes can be solved by an iterative method using
Green's functions \cite{Gong:2001he}. To second order in the slow-roll
parameters, here expressed as horizon-flow parameters, the seed for recursion
(\ref{eq:an}) for the scalar spectrum reads
\begin{eqnarray}
\label{eq:scalarSG}
a_{{\rm S}0} &=& 1 - 2(C+1) \epsilon_1 - C \epsilon_2 \nonumber \\
& & + \left(2C^2 + 2C + \frac{\pi^2}2 - 5\right) \epsilon_1^2
+ \left(\frac{C^2}2 + \frac{\pi^2}8 - 1\right) \epsilon_2^2 \nonumber \\
& & + \left(C^2 - C + \frac{7 \pi^2}{12} - 7\right)
\epsilon_1\epsilon_2 + \left(- \frac{C^2}2 +
\frac{\pi^2}{24}\right) \epsilon_2\epsilon_3\, ,
\end{eqnarray}
where $C \equiv \gamma_{\rm E} + \ln 2 - 2\approx -0.7296$,
while for the tensor spectrum we have \cite{Leach:2002ar},
\begin{eqnarray}
a_{{\rm T}0}  &=& 1 - 2(C+1) \epsilon_1 \nonumber \\
& & + \left(2C^2 + 2C + \frac{\pi^2}2 - 5\right) \epsilon_1^2 +
\left(-C^2 - 2C + \frac{\pi^2}{12} - 2\right) \epsilon_1\epsilon_2\, .
\end{eqnarray}
In Fig.~\ref{fig:fig1} we estimated the region of parameter space in which
the prediction of the pivot amplitude $A(k_*)$ is better than $0.3\%$ for
the slow-roll approximation at second order. Such a high precision is actually
needed to ensure that the power spectrum can be predicted with an accuracy
better than a few per cent over at least three or four decades in wave number.

\subsubsection{Constant-horizon assumption.}
\label{sssec:cha}

In a number of inflationary models the time derivative of the Hubble
distance is tiny (see Section \ref{sec:test} for examples). For this
kind of models, during a certain number of e-foldings, $\epsilon_1
\ll 1$. However, as we have seen in Section~\ref{sec:hubble}, this
does not necessarily mean that all other $\epsilon_m$ have to be
small as well. We thus worked out the constant-horizon approximation
\cite{Schwarz:2001vv} at order $q$ for the situation $|\epsilon_2^q|
> \max(|\epsilon_1\epsilon_2|, |\epsilon_2\epsilon_3|)$, which means
that, for $a_0$ we are allowed to include the following monomials in
the primordial spectra: $1, \epsilon_1, \epsilon_2, \dots,
\epsilon_2^q$. With this approximation, the $q=3$ expression for the
seed of the scalar spectrum is
\begin{eqnarray}
\label{eq:scalarCHA}
a_{{\rm S}0} &=& 1 - 2(C+1) \epsilon_1 - C \epsilon_2
        + \left(\frac{C^2}2 + \frac{\pi^2}8 - 1\right)\epsilon_2^2 \nonumber \\
    & & - \left(\frac{C^3}6 - C + C\frac{\pi^2}8 + \frac 78 \zeta(3)
        - \frac 23 \right)\epsilon_2^3 \, ,
\end{eqnarray}
where $\zeta(3) \approx 1.2021$, and to any order for the tensor
spectrum,
\begin{eqnarray}
a_{{\rm T}0}  &=& 1 - 2(C+1) \epsilon_1 \, .
\end{eqnarray}
Also here we estimate the region of $0.3\%$ accuracy of $A(k_*)$ and indicate
it in Fig.~\ref{fig:fig1}. For more details on how to do this we refer
the reader to the work of \cite{Schwarz:2001vv}.

\subsubsection{Growing-horizon assumption.}

This approximation is valid for cases with $|\epsilon_m| <
\epsilon_1$ for $m > 1$. The trivial example here is power-law
inflation \cite{Lucchin:1984yf}, where only $\epsilon_1 \neq 0$.
This approximation is also valid in inflationary scenarios, where
power-law inflation dynamics is a past or future attractor, and
$\epsilon_1$ is sufficiently large for higher-order corrections to
make sense. An expression for $a_0$ can be obtained by keeping all
terms in $\epsilon_1$ up to order $q$, where $q$ is the maximal
integer for which $\epsilon_1^q > \max(|\epsilon_1
\epsilon_2|,|\epsilon_2 \epsilon_3|)$ holds true. We defined the
(linearly) growing-horizon approximation to order $q$
\cite{Schwarz:2001vv} to include the following terms:
$1,\epsilon_1, \dots, \epsilon_1^q, \epsilon_2$. To third order
the corresponding expression for the scalar spectrum is
\begin{eqnarray}
a_{{\rm S}0} &=& 1 - 2(C+1) \epsilon_1
        + \left(2C^2 + 2C + \frac{\pi^2}{2} - 5\right)\epsilon_1^2  \nonumber \\
    & & -\ \left(\frac{4C^3}3 + C\pi^2 - 12C + \frac{14}3 \zeta(3)
        - \frac{19}3\right)
          \epsilon_1^3 - C \epsilon_2 \, ,
\end{eqnarray}
and for the tensor spectrum,
\begin{eqnarray}
a_{{\rm T}0}  &=& 1 - 2(C+1) \epsilon_1 \nonumber \\
    & & + \left(2C^2 + 2C + \frac{\pi^2}2 - 5\right) \epsilon_1^2 \nonumber \\
    & & -\ \left( \frac{4C^3}3 + C\pi^2 - 12C + \frac{14}3 \zeta(3)
        - \frac{19}3\right) \epsilon_1^3\, .
\end{eqnarray}

\section{Classification of inflation models in the $r$--tilt plane}

In this section we indicate how to make contact with the variables that are frequently used
in cosmological parameter estimation.
However useful these variables have been assumed to be so far, as we shall see here,
the analysis using them
becomes much more involved that while using the horizon-flow functions.
Complications arise because there are not exact model independent expressions for
$(n-1)$ and $r$; they are given as expansions in terms of the horizon-flow functions.

Using one or the other set of variables is just a matter of choosing
the working parametrization for the primordial spectra. Very often a
power-law shape is assumed for the scalar power spectrum;
\begin{equation}
{\cal P}_{\rm S}(k) = A(k_*) \left(k\over k_*\right)^{n-1},
\end{equation}
where $n$ is called the spectral index and $n-1$ the tilt of the
spectrum. A comparison with expression (\ref{eq:power}) reveals that
\begin{eqnarray}
A(k_*) &=& \tilde{{\cal P}}(k_*) a_{{\rm S}0}, \\
(n-1)(k_*) &=& {a_{{\rm S}1} \over a_{{\rm S}0}}.
\end{eqnarray}
For the slow-roll and the constant-horizon approximation we find
at second order in the horizon-flow functions
\begin{equation}
(n-1)(k_*) = -2 \epsilon_1 - \epsilon_2
      - 2 \epsilon_1^2 - (2C+3)\epsilon_1\epsilon_2 - C \epsilon_2\epsilon_3.
\end{equation}
A difference between the two approximations shows up at the third order. Here
we restrict the discussion to the second order expressions.

Tensor contributions are typically introduced via the tensor to scalar ratio
\begin{equation}
r = {{\cal P}_{\rm T}(k_*)\over {\cal P}_{\rm S}(k_*)}
= 16 \epsilon_1 (1 + C \epsilon_2)
\end{equation}
at second order in the slow-roll and the constant-horizon approximation.

\begin{figure}
\centerline{\includegraphics[width=0.6\linewidth]{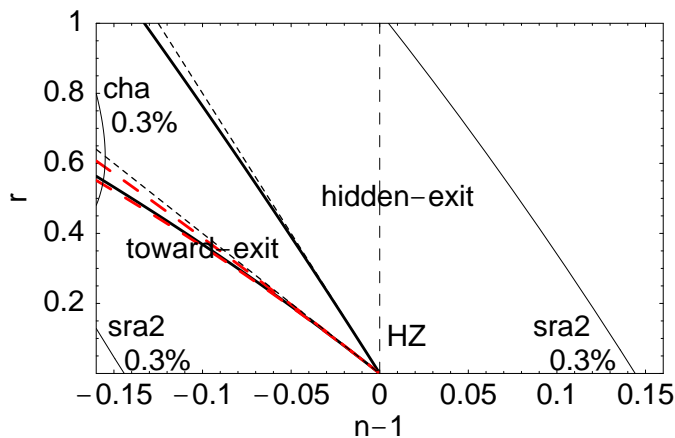}}
\caption{Classification of models in the $r$--tilt plane. The vertical long
dashed line denotes the scale-invariant HZ spectrum. Toward-exit models
of inflation have a negative tilt (red spectrum) and are found
to the l.h.s. of the upper full line.
To the r.h.s. of the upper full
line we find the hidden-exit models. The (upper) short dashed line
shows the approximate borderline derived from the leading order expression
given in the text. The toward-exit models
are divided by another thick line which assumes vanishing running of
the spectral index. Models with decreasing kinetic energy
are confined between both thick lines.
The (lower) short dashed line is the approximate
leading order result. The tilted long dashed lines show two cases of
non-vanishing running of the spectral index, namely ${\rm d}n/{\rm
d}\ln k = -12 \epsilon_1^2 (+4 \epsilon_1^2)$, corresponding to
$\epsilon_3 = + 2 \epsilon_2 (- 2 \epsilon_2)$. The upper curve
corresponds to the negative sign of the running. We also indicate
the accuracy of the considered approximations. } \label{fig:nr}
\end{figure}

A closer inspection of the above expressions shows that there is a simple
one-to-one map $(\epsilon_1,\epsilon_2) \leftrightarrow (r,n-1)$ at the
leading order, namely $\epsilon_1 = r/16$ and $\epsilon_2 = -r/8 - (n-1)$.
Thus the physical classification of single scalar field models is characterized
by the borderlines $\epsilon_2 = 0 \Leftrightarrow r = - 8 (n-1)$ and
$\epsilon_2 = \epsilon_1/2  \Leftrightarrow r = - 4 (n-1)$ (short dashed lines
in Fig.~\ref{fig:nr}).

This one-to-one correspondence is spoiled by the $\epsilon_3$-dependence
of $n-1$ at the second order. Nevertheless, the borderline $\epsilon_2 = 0$
(upper full line in Fig.~\ref{fig:nr}) is well defined, since higher horizon-flow
parameters enter only together with $\epsilon_2$ and thus it is in principle
possible to distinguish toward-exit inflation from hidden-exit inflation
in the $r$-tilt plane.

However, the borderline ($\epsilon_2 = \epsilon_1/2$) between the two subclasses of
toward-exit models, i.e., models with general
initial conditions and models with special initial conditions,
gives rise to a family of lines that
depends on the observed running of the spectral index,
${\rm d}n/{\rm d}\ln k = - 2\epsilon_1\epsilon_2 - \epsilon_2\epsilon_3$
(leading order). One possibility is to fix $\epsilon_3$ by assuming that
the leading contribution to the running vanishes (lower full line in
Fig.~\ref{fig:nr}). Thus it is impossible to distinguish between models with
decreasing (``chaotic'') and increasing (``false vacuum'') kinetic energy
density on the basis of a $r$--tilt plot, unless the running of the spectral
index is known or constrained to be small.

The above discussion confirms that the horizon-flow functions are more
fundamental than other quantities that have been used to classify inflation
models. However, the upper solid line in Fig.~\ref{fig:nr} is robust in the
sense that it holds true for the slow-roll approximation and the
constant-horizon approximation at any order (as terms $\epsilon_n$ with $n >2$
enter in combination with $\epsilon_2$ and are thus zero for $\epsilon_2 = 0$).

Figure \ref{fig:nr} shows another remarkable feature of the constant and
growing horizon approximations: for the shown region in $r$ and $n-1$ the
amplitudes are accurate to $0.3\%$, except for the tiny region enclosed by
the arc on the l.h.s. of the figure. This is in contrast to the slow-roll
approximation at second order, which is less accurate in the upper right
and lower left corner of the figure.

\section{Testing the higher-order constant-horizon approximation}
\label{sec:test}

As we already noted, observations show that $\epsilon_1 \ll
1$. If $\epsilon_2$ is of the order of $\epsilon_1$ or smaller,
higher-order corrections (terms that are at least quadratic in the
horizon-flow functions) to the primordial spectra are irrelevant
and the first-order expression for $a_0$, the seed of the
recursion (\ref{eq:an}), is good enough to calculate the
coefficients $a_n$ in the power spectrum (\ref{eq:power}).
Nevertheless, if $\epsilon_2$ is actually much larger than
$\epsilon_1$, then higher-order corrections in $\epsilon_2$ could be
necessary to match the observational accuracy. In such a case the
constant-horizon approximation (see Section \ref{sssec:cha}) is the
simplest and most economic way of obtaining high-precision predictions. To see
if this is true, let us start by comparing Eqs.~(\ref{eq:scalarSG}) and
(\ref{eq:scalarCHA}). One sees that the third-order expression for the
constant-horizon approximation is simpler and requires knowledge of
less horizon-flow functions than the corresponding lower-order slow-roll
approximation. Next, we must test how good the constant-horizon
approximation is.

To measure the error of the approximations, we define
\begin{equation}
\label{eq:error} \Delta A(\log \tilde k)= \frac{A_{\rm num}(\log
\tilde k) - A_{\rm appr}(\log \tilde k)} {A_{\rm num}(\log\tilde
k)}\, 100\% \, ,
\end{equation}
where $A_{\rm num}(\log \tilde k)$ and $A_{\rm appr}(\log \tilde
k)$ stand for the numerical and analytically approximated values
of the amplitudes at the normalized scale $\tilde k\equiv k/k_*$.

The constant-horizon approximation applies for many single-field
models based on phenomenological particle physics: the inverted quadratic
model $V = V_0 - m^2 \varphi^2/2$ \cite{Binetruy:ss}, or more generally
models with $V = V_0[1 - (\varphi/\mu)^p]$ and $p\geq 2$, and also for
those with $V = V_0(1 - \exp(-\varphi/\mu))$ \cite{Stewart:1994ts}. For
positive $\mu$, all these models belong to the class of toward-exit inflation
with special initial conditions. There are also examples of hybrid
inflation models, e.g.\ those arising from dynamical supersymmetry
breaking with $V = V_0[1 \pm (\mu/\varphi)^p]$, where $p$ is a
positive integer \cite{Kinney:1998dv}. For the positive sign, the model
belongs to the hidden-exit inflation class, while for the negative sign, the
models belong to the class of toward-exit inflation with special initial
conditions. Yet another model belonging to that class is hybrid inflation
with a running mass that arises from one-loop corrections in
supersymmetry-inspired models
(see Refs.~\cite{Stewart:1996ey,Stewart:1997wg,Covi:1998yr}
for concrete realizations):
\begin{equation}
\label{eq:rmhiV2} V = V_0\left[1 - \frac12 m^2(\varphi)\varphi^2\right] \, .
\end{equation}
Further examples where the constant-horizon approximation applies
are the so-called natural inflation model \cite{Freese:1990rb}:
\begin{equation}
\label{eq:ni} V = \Lambda^4\left[1+\cos\left(N\frac{\varphi}{f}\right)\right]
\, ,
\end{equation}
and the model proposed by Wang et al. \cite{Wang:1997cw},
\begin{equation}
\label{eq:WangV} V =
\Lambda^4\left[1-\frac2\pi\arctan\left(5\varphi\right)\right]
\, ,
\end{equation}
which is a designer model that has been used to demonstrate the limitations
of the slow-roll approximation.

We started by testing the natural inflation model given by
potential (\ref{eq:ni}). For this model with $\Lambda=f=1$ and
$N=1$ at the pivot value $\varphi_*=0.01$, we obtained
$H_*=0.81648$, $\epsilon_1 \approx \epsilon_3 \approx 10^{-7}$,
$\epsilon_2 \approx 0.039529$ and $\epsilon_4 \approx 0.02008$.
With these values, we find that the third-order constant-horizon
approximation performs slightly better than the second-order slow-roll
approximation, and that both of them provide a significant improvement with
respect to the first order slow-roll approximation. The errors are confined
within the $15\%$ interval in a range exceeding $\Delta \log \tilde k =6$.

\begin{figure}
\centerline{\includegraphics[width=0.55\linewidth]{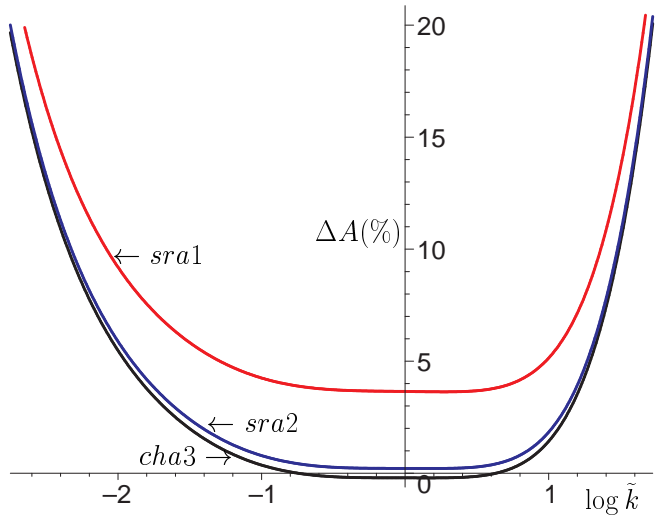}}
\caption{Test of approximated solutions to the equation of the
scalar modes for a hybrid model with running mass. Third-order
constant-horizon approximation (cha3), and first (sra1) and second
(sra2) order slow-roll approximations are compared. The errors shown
are those obtained including terms up to $a_3$ in parametrization
(\ref{eq:power}) and keeping the highest possible order for the
horizon-flow functions in all coefficients.} \label{fig:hrm1}
\end{figure}

\begin{figure}
\centerline{\includegraphics[width=0.55\linewidth]{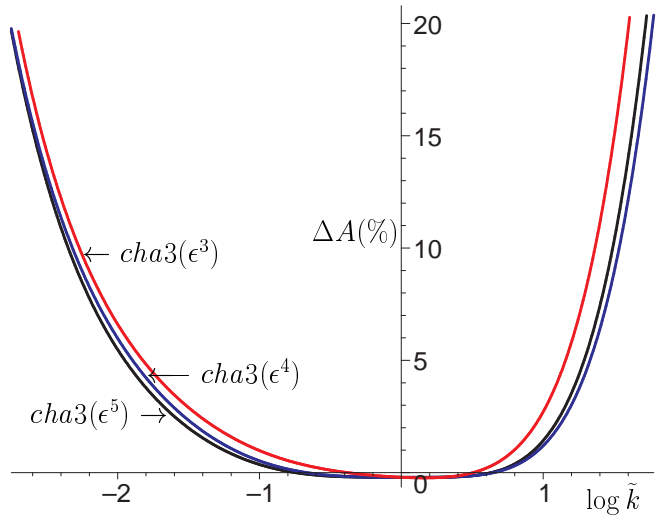}}
\caption{Test of the constant-horizon approximation for a hybrid
model with running mass. The errors shown are those obtained
including terms up to $a_3$ in parametrization (\ref{eq:power}), and
varying the order of the highest included horizon-flow function.}
\label{fig:hrm2}
\end{figure}

\begin{figure}
\centerline{\includegraphics[width=0.55\linewidth]{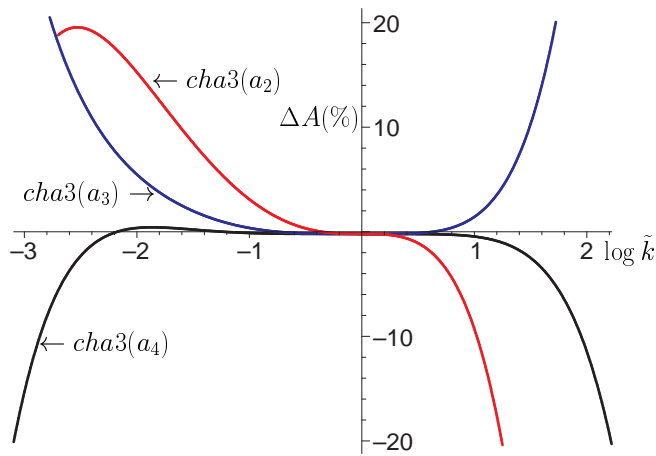}}
\caption{Test of the constant-horizon approximation for a hybrid
model with running mass. Here we investigate the convergence of the
Taylor expansion (\ref{eq:power}) as more and more coefficients
$a_n$ are added. We keep the highest possible order of the
horizon-flow functions.} \label{fig:hrm3}
\end{figure}

More interesting is a test for a model where some of the higher horizon-flow
functions are ``large''. We start from Eq.~(\ref{eq:rmhiV2}) and use the
ad hoc choice $m^2(\varphi)=M^2 \exp\left(-\varphi/\mu\right)$, with $M$ and
$\mu$ positive. The resulting potential has a maximum at $\varphi=0$ and a
minimum at $\varphi=2 \mu$. Starting near the false vacuum, enough inflation
can be produced before the minimum is reached. For $V_0=1$,
$\mu=1$ and $M=2$ at the pivot value $\varphi_*=0.01$, we find
$H_*=0.57729$, $\epsilon_1=0.00003$, $\epsilon_2=0.29494$,
$\epsilon_3=-0.00381$, $\epsilon_4=0.14344$ and
$\epsilon_5=-0.0082$, with the results presented in
Figs.~\ref{fig:hrm1}, \ref{fig:hrm2} and \ref{fig:hrm3}.

As can be observed in Fig.~\ref{fig:hrm1}, the conclusions drawn
for the simple inflation near a maximum model are still valid for
models where the conditions for constant-horizon approximation to
apply are met in a weaker fashion than in the case of natural
inflation, although the range where the errors are confined in the $15\%$
error band is smaller. According with the results in
Fig.~\ref{fig:hrm2}, increasing the order of the horizon-flow
functions beyond the quadratic in every included term of parametrization
(\ref{eq:power}) does not significantly improve the accuracy of
the approximation.

As shown in Fig.~\ref{fig:hrm3}, adding terms in parametrization
(\ref{eq:power}) actually seems to be most important for increasing the
precision of the prediction, but this conclusion might not hold true for
other models. Even in the best case, it seems difficult to keep the
error below the $15\%$ mark for a range broader that $\Delta \log \tilde k
= 5$.

We tried to push our approximation to the limits by testing it on the model
given by Eq.~(\ref{eq:WangV}). For $\Lambda=1$ and $\varphi_*= - 0.25$,
we find $H_*=0.72476$, $\epsilon_1=0.01122$,
$\epsilon_2=0.31874$, $\epsilon_3=0.19659$, $\epsilon_4=0.079$ and
$\epsilon_5=0.129$. The test confirms the previous
results, although the range where the error is
under $15\%$ is significantly shorter, $\Delta \log \tilde k
\approx 3.5$.

For completeness we note here that similar results were obtained for the
tensor modes.

For all tested cases, the constant-horizon approximation at third order
performed as good or slightly better than the slow-roll approximation
at second order, but it has the advantage that one needs to make less
restrictive assumptions on higher horizon-flow functions and the approximation
is more efficient in the sense that the recursion seed $a_0$ only needs
$\epsilon_1$ and $\epsilon_2$ as an input, whereas the slow-roll approximation
at second order needs additionally $\epsilon_3$.

\section*{Acknowledgements}

We thank Samuel Leach, Andrew Liddle, J\'{e}r\^{o}me Martin and
Slava Mukhanov for discussions and Laura Covi for
references. We are particularly grateful to
Samuel Leach for providing the data for the
$\epsilon_1$--$\epsilon_2$ plane contours and the algorithm for the
numerical solution of the equation of perturbations. CTE
acknowledges the hospitality of CERN TH and the partial
support by CONACyT (38495--E) and SNI from Mexico.

\end{document}